\documentclass[journal,comsoc]{IEEEtran}
\IEEEoverridecommandlockouts
\usepackage[numbers]{natbib}
\setcitestyle{open={[},close={]}}

\usepackage{amsmath,amssymb,amsfonts}
\usepackage{graphicx}
\usepackage{textcomp}
\usepackage{xcolor}
\usepackage{float}
\usepackage{array}
\usepackage[font=footnotesize,labelfont=bf]{caption}
\usepackage[utf8]{inputenc}
\def\BibTeX{{\rm B\kern-.05em{\sc i\kern-.025em b}\kern-.08em
    T\kern-.1667em\lower.7ex\hbox{E}\kern-.125emX}}
\usepackage{algorithm}
\usepackage{algorithmic}

\usepackage{atbegshi}

\begin{document}

\title{Federated Edge Learning : Design Issues and Challenges\\}

\author{
Afaf Taïk \IEEEmembership{Student Member, IEEE} and Soumaya Cherkaoui, \IEEEmembership{Senior Member, IEEE}

\thanks{ Afaf Taïk and Soumaya Cherkaoui are with the INTERLAB Research Laboratory, Faculty of Engineering, Department of Electrical and Computer Science Engineering, University of Sherbrooke, Sherbrooke, QC J1k 2R1, Canada,  (e-mail: afaf.taik@usherbrooke.ca, soumaya.cherkaoui@usherbrooke.ca).}

}
\maketitle

\begin{abstract}
Federated Learning (FL) is a distributed machine learning technique, where each device contributes to the learning model by independently computing the gradient based on its local training data. It has recently become a hot research topic, as it promises several benefits related to data privacy and scalability. However, implementing FL at the network edge is challenging due to system and data heterogeneity and resources constraints.
In this article, we examine the existing challenges and trade-offs in Federated Edge Learning (FEEL).  The design of  FEEL algorithms  for resources-efficient learning raises several challenges. These challenges are essentially related to the multidisciplinary nature of the problem. As the data is the key component of the learning,  this  article  advocates a new set of considerations for data characteristics in wireless scheduling algorithms in FEEL. Hence, we propose a general framework for the data-aware scheduling as a guideline for future research directions. We also discuss the main axes and requirements for data evaluation and some exploitable techniques and metrics.   
\end{abstract}

\IEEEpeerreviewmaketitle


\section{Introduction}
\label{sec:introduction}
The growing interest in intelligent services motivates the integration of artificial intelligence (AI) in Internet of Things (IoT) applications. The collection of large volumes from the different devices and sensors is necessary  for  training AI  models. 
However, uploading massive data generated by connected devices to the cloud is usually impractical, mainly due to issues including privacy, network congestion, and latency. 
Federated Edge Learning (FEEL) \cite{r2} is a Machine Learning (ML) setting that utilizes edge computing \cite{filali_multi-access_2020,abouaomar_resource_2021} to tackle these concerns. In contrast to centralized ML, Federated Learning (FL) \cite{konecny_federated_2017} consists of training the model on the devices, with the orchestration of a central entity, where only the resultant model parameters are sent to the edge servers to be aggregated.
FEEL refers to the use of FL at the edge of the network, which makes it a promising solution for privacy preserving ML. 

An important design decision for a FEEL algorithm is whether to choose either asynchronous or synchronous aggregation. Recent works tend to promote synchronous training, where, for instance, synchronization among participating devices is required for updates averaging \cite{konecny_federated_2017} and privacy-preservation \cite{wei_federated_2019}. However, there are many challenges upon using synchronous FL in edge environments.

To begin with, the heterogeneity of resources across different devices sparks new system challenges. For instance, significant delays can be caused by stragglers. Moreover, communication loads across devices limit the scalability of FL for large models. Participating devices communicate full model updates during every training iteration, which are of the same size as the trained model. For large models, such as deep neural networks, the model size can be in the range of gigabytes. As a result, if communication bandwidth is limited or communication is costly, FEEL can be deemed impractical or unfeasible, as communication overhead becomes a bottleneck for FEEL. 

Furthermore, end devices have limited battery lives and varying available energy levels. As training ML models is a computation-heavy task, only devices that have enough energy can be solicited to participate. Furthermore, energy and computational constraints limit both the size of the models that can be trained on-device, and the number of local training iterations.         

Additionally, as the data  collected by the clients depends on their local environment and usage pattern, both the size and the distribution of the local datasets will typically vary between different clients. This non-Independently and Identically Distributed (non-IID) and unbalanced nature of data across the network imposes significant challenges linked to models' convergence.

Consequently, designing an efficient FEEL algorithm  should  take into account the limited and heterogeneous nature of the resources, alongside the non-IID and unbalanced aspect of the data distributions. In general, proposed FEEL algorithms target efficient selection of participant devices, optimization of the resource allocation and usage, or adequate updates' aggregation. However, it is hard to capture both the resources problems and the learning goal, as there is no direct relation between the model’s loss function and the resource optimization. A manageable approach found in current works is to focus on resource optimization with certain learning guarantees, such as maximizing the number of collected updates and maintaining the level of local accuracy \cite{zeng_energy-efficient_2019}. Nonetheless, these guarantees are not sufficient, as a significant drop in accuracy is observed when data is non-IID and unbalanced. 
Therefore, we propose to lighten the effects of design trade-offs through the direct integration of the data properties in the device selection and resource optimization algorithms. In fact, data properties were at the heart of FL since its inception, but they have been largely overlooked in the design of FEEL algorithms. Moreover, data diversity has long been premised on in active learning, where models can be trained using few labelled data samples if the highly diverse data is selectively added to the training set. Thus, data diversity should  be considered in  the  design  of FEEL algorithms, as we advocate in this article.

The main contributions of this article can be summarized as follows:
\begin{itemize} 
\item We discuss the FEEL challenges imposed by the nature of the edge environment, from an algorithms design perspective. We review the challenges related to computational and communication capacities, as well as data properties, as they are at the core of the trade-offs in learning and resource optimization algorithms. 
\item We  propose a general framework for incorporating data properties in FEEL, by providing a guideline for a thorough algorithm design, and criteria for the choice of diversity measures in both datasets and models.  
\item We present several possible measures and techniques to evaluate data and model diversity, which can be applied in different scenarios (e.g., classification, time series forecasting), in an effort to assist fellow researchers to further address FEEL challenges.
\end{itemize}

The remainder of this article is as follows. In Section II, we review the challenges found in designing FEEL algorithms, and we derive the main trade-offs. Then, we shed the light on a new data-aware design direction for FEEL algorithms in section III. Some possible techniques and methods to evaluate diversity are detailed in this section. At last, a conclusion and final remarks are presented in Section IV.

\section{Design challenges : Overview}
\label{sec:Challenges}
FEEL has several constraints related to the nature of the edge environment. In fact, FEEL involves the participation of heterogeneous devices that have  different computation and communication capabilities, energy  states, and dataset  characteristics. Under device and data heterogeneity, in addition to  resources  constraints,  participants selection \cite{fedcs} and resource allocation \cite{resource_amine} have to  be  optimized  for an efficient FEEL solution.

\subsection{Design Challenges}
The  core  challenges  associated  with  solving  the  distributed  optimization  problem are twofold: Resources and Data.  These  challenges  increase the  FEEL setting  complexity compared to similar problems,  such  as  distributed  learning  in  data  centers. 

\textbf{Resources:} The challenges related to the resources, namely computation, storage and communication, are mainly in terms of their heterogeneity and scarcity. 

\textit{Heterogeneity of the resources:} The computation, storage and communication capabilities vary from a device to another. Devices may be equipped with different hardware (CPU and memory), network connectivity (e.g., 4G/5G, Wi-Fi),  and may differ in available power (battery  level). 
The gap in computational resources creates challenges such as delays caused by stragglers. FEEL algorithms must therefore be adaptive to the heterogeneous hardware and be tolerant toward device drop-out and low or partial participation. A potential solution to the straggler problem is asynchronous learning. However, the reliability  of asynchronous FL and the model convergence in this setting are not always guaranteed. Thus,  synchronous FL remains the preferred approach.
 
\textit{Limited Resources:} In a contrast to the cloud, the computing and storage resources of the devices are very limited. Therefore the models that can be trained on device are relatively simpler and smaller than the models trained on the cloud. Furthermore, devices are frequently offline or unavailable either due to low battery levels, or because their resources are fully or partially used by other applications. 

As for the communication resources, the available bandwidth is limited. It  is  therefore  important  to  develop  communication-efficient  methods that allow to send  compressed or partial  model  updates. To further reduce communication cost in FEEL settings, two potential directions are generally considered 1) reducing the total number of communication rounds until convergence \cite{wang_cmfl_2019}, and 2) reducing the size of the transmitted updates through compression and partial updates \cite{konecny_federated_2017}.

\textbf{Data:} 
In most cases, data distributions depend on the users' behaviour \cite{self}. As a result,  the local datasets are massively distributed, statistically heterogeneous (i.e., non-IID and unbalanced), and highly redundant. Additionally, the raw generated data is often privacy-sensitive as it can reveal personal and confidential information.      

\textit{Small and widely distributed datasets:} In FEEL scenarios, a large number  of devices  participate in  the FL training with a small average number of data samples per client. Learning from small datasets makes local models prone to overfitting.  

\textit{Non-IID:} The training data on a given device is typically based on the usage of the device by a particular user, and hence any particular user’s local dataset will not be representative of the population distribution. This data-generation paradigm fails to comply with the independent and identically distributed (IID) assumptions in distributed optimization, and thus adds complexity to the problem formulation and convergence analysis. The empirical evaluation of FEEL algorithms on non-IID data is usually performed on artificial partitions of MNIST or CIFAR-10, which do not provide a realistic model of a federated scenario.  

\textit{Unbalance:} Similarly to the nature of the distributions, the size of the generated data depends on the user. Depending on users' use of the device, these may have varying amounts of local training data. 

\textit{Redundancy:} The unbalance of the data is also observed within the local datasets at a single device. In fact, IoT data is highly redundant. In sequential data (e.g., video surveillance, sensors data) for instance, only a subset of the data is informative or useful for the training.   

\textit{Privacy:} 
The privacy-preserving aspect is an essential requirement in FL applications. The raw data generated on each device is protected by sharing only model updates instead  of  the  raw  data.  However,  communicating  model  updates  throughout  the  training  process  can still be reverse-engineered to reveal sensitive information,  either  by  a  third-party or a malicious central server.

\subsection{Design Trade-offs}
Several efforts were made to tackle the aforementioned challenges. However,  FEEL is a multi-dimensional problem that brings about several trade-offs. As a result, algorithms designed to address one issue at a time are deemed unpractical.
Perhaps a tractable solution may be to combine several techniques when developing and deploying FEEL algorithms.

In general, an end-to-end FEEL solution should cover devices selection, resource allocation, and updates aggregation. In the following, we discuss major trade-offs  that should be considered when designing solutions in the FEEL setting.\\
\newline
\textit{1) General FEEL solution}
\newline
Given the wide range of applications that can benefit from FEEL, there is no one-size-fits-all solution. However, in general, a FEEL solution needs to act on the following aspects: \\

\textbf{Device selection:} Participant  selection  refers  to  the selection of devices to receive and train the model in each training round. Ideally, a set of participants is randomly selected by the server  to  participate.  Then,  the  server  has  to  aggregate parameter  updates  from  all  participants  in  the  round before  taking  a  weighted  average  of  the  models. However, due to the communication bottlenecks and the desire to tame the training latency, the device selection should be optimized in terms of resources \cite{fedcs} and data criteria. 

\textbf{Resource allocation:} Device selection should not be considered independently from resource allocation, especially computation and bandwidth. We refer to the joint selection and resource allocation as a scheduling algorithm.  Indeed, the number of scheduled devices is limited by the available bandwidth that can be allocated. Additionally, for an optimal learning round duration and energy consumption, both bandwidth and computation resources should be adapted based on the number of local iterations at each device, and the number of global iterations (i.e., learning rounds) \cite{wang_adaptive_2019}. 
Due to the fast-changing aspect of the FEEL environment, the computational complexity of scheduling algorithms should be especially low. Therefore, the use of meta-heuristics and heuristics should be encouraged.

\textbf{Updates aggregation:} This aspect of the solution design refers to how the updates are aggregated and how frequently they are aggregated. 
For instance, the frequency of the communication and aggregation can be reduced with more local computation \cite{wang_adaptive_2019}, or reduced through selective communication of gradients \cite{wang_cmfl_2019}.
For instance, FedAvg \cite{zhao_federated_2018} is one of the most used methods in aggregation which uses weighted average Stochastic Gradient Descent updates, where the corresponding weights are decided by the volume of the training dataset. 
While FedAvg uses synchronous aggregation, in FedAsync \cite{xie_asynchronous_2019} algorithm, newly  received  local  updates  are weighted  according  to  their staleness, where stale updates received from stragglers are weighted less based on how many rounds elapsed. 
It should also be noted that proposing new aggregation methods requires theoretical and empirical convergence analysis to guarantee that the learning loss function will converge to a global optimum.  Updates aggregation should also be communication-efficient \cite{konecny_federated_2017, wang_cmfl_2019} and secure by the means of techniques such as differential privacy \cite{wei_federated_2019}. \\
\newline
\textit{2) Optimization axes}
\newline
\begin{figure*}
	\centering
	\includegraphics[scale=0.52]{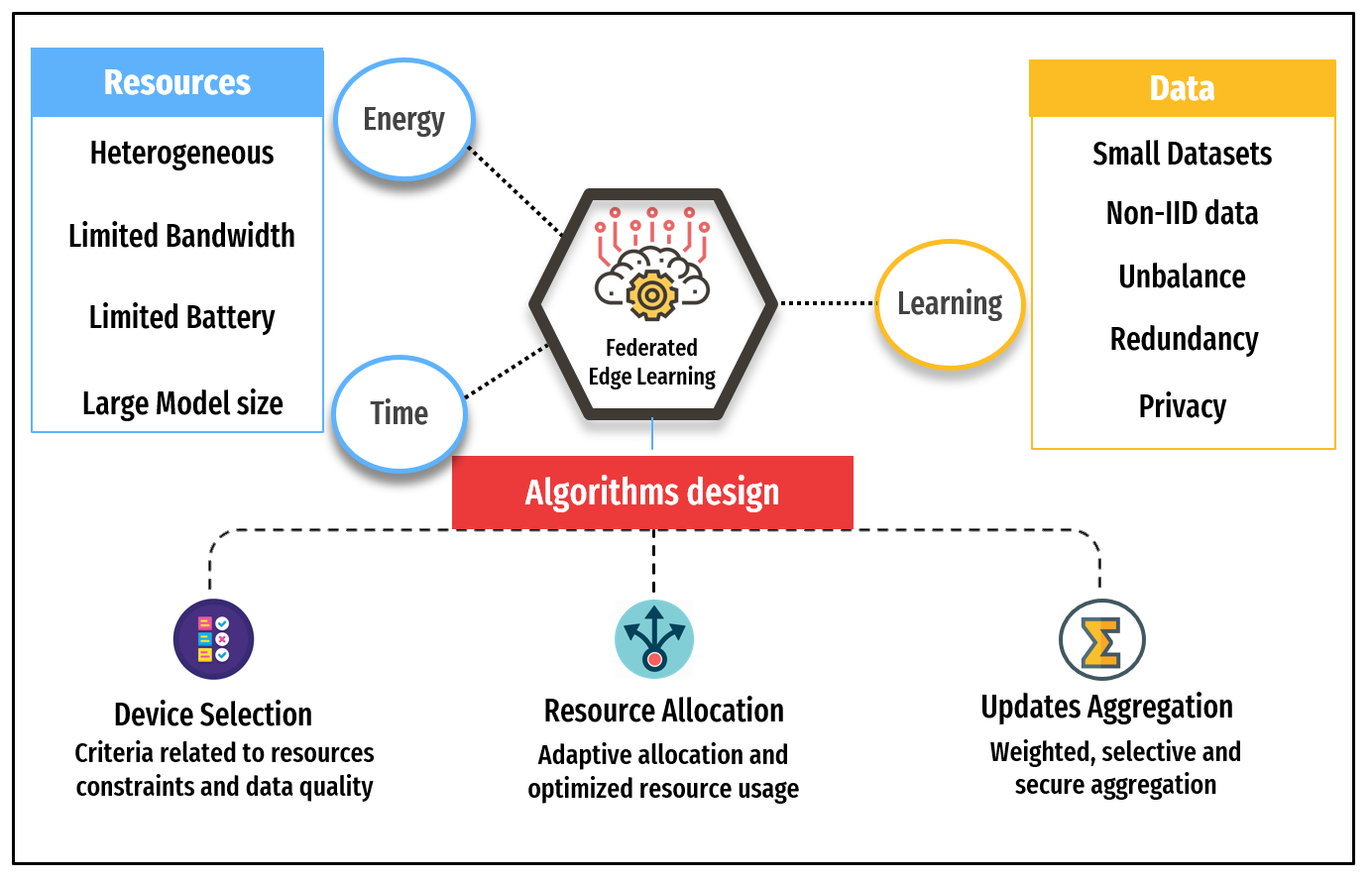}
	\caption{FEEL algorithms, challenges and optimization axes}
	\label{fig:fig_algos}
\end{figure*}

In addressing FEEL challenges, three optimization axes are often considered: Time, Energy and Learning. In many cases, the FEEL algorithm can be viewed as a Pareto optimal problem \cite{reviewer2}.  The relation between the three axes and the challenges is illustrated in Figure \ref{fig:fig_algos}. 

\textbf{Time optimization:} Accelerating the learning time can be evaluated with different lenses: learning round duration and time until learning convergence. 
Due to the synchronous model aggregation of FEEL, the total duration of a  round is determined by the slowest device among all the scheduled devices \cite{shi_device_2019}. For this reason,  more bandwidth should be allocated for transmission by stragglers and less for faster devices. This to some extent can equalize their total update time (computing plus communication time).  Furthermore, to avoid squandering bandwidth on extremely slow devices, scheduling (i.e., joint selection and resource allocation) should exclude slowest devices by applying thresholds on their expected completion time, which can be inferred using their computing capacities and channel states. From a learning perspective, the learning latency is determined by the number of rounds until convergence. The optimization techniques centered on this aspect mainly focus either on the selective upload of updates, or on maximizing the participating devices in each round.

\textbf{Energy optimization:} Optimizing the energy consumption across the network is necessary to reduce the rate of drop-out devices because of battery drainage. In fact, training and transmission of large-scale models are  energy  consuming,  while  most  edge and end devices  have  limited  battery  lives. Additionally, using the maximum capacity of the devices would make the users less-likely willing to participate in the training.  
A design goal of a scheduling algorithm (i.e., joint selection and resource allocation) would be to allocate bandwidth based on the devices' channel states and battery levels. As a result, more bandwidth should be allocated to devices with weaker channels or poorer power states, to maximize the collected updates \cite{zeng_energy-efficient_2019}.

\textbf{Learning optimization:} 
In contrast to centralized learning, optimizing the learning in the FEEL setting cannot be seen independently from time and energy optimization.  
However, capturing the optimization of time, energy and the learning goal in the same optimization problem is hard, because there is no direct relation between the objective function of the learning (i.e., the loss function) and the time and energy minimization goal. A manageable approach used is to minimize time and energy under a certain convergence speed guarantee.
For instance, some works argue that the number of collected updates in each round is inversely proportional to the convergence speed, and therefore is used as a guarantee \cite{zeng_energy-efficient_2019}. 
Indeed, multi-user diversity (i.e., collecting a maximum of updates) can yield a high convergence speed, especially in IID environments, however there is a significant chance of choosing the same sets of devices repeatedly. To avoid this issue, a goal of the FEEL algorithm can be to maximize the fairness in terms of the number of collected updates among devices \cite{yang_age-based_2020}. The fairness measure maximizes the chance of more diverse data sources, thus achieving gradient diversity.
Nonetheless, the number of collected updates in this setting might be low. The fairness is also considered in the aggregation by q-Fair FL (q-FFL) \cite{li_fair_2020}, which reweighs the objective function in FedAvg to assign higher weights  in  the  loss  function  to  devices  with higher  loss. 
Another approach is to use  data size priority, which maximizes the size of data used in the training,  by using a probability of selection inversely proportional to the available dataset's size. 
In the background, these scheduling algorithms all share the same idea : if the size of the training data is large then the training would converge faster. However, IoT data is highly redundant and inherently unbalanced. Thus, many of the proposed algorithms witness a drop in performance in non-IID and unbalanced experiments. Therefore, the data properties should be considered throughout the FEEL algorithm.

\section{Data-aware FEEL design: Future Direction}
\label{sec:FEEL}
\begin{figure*}
	\centering
	\includegraphics[scale=0.50]{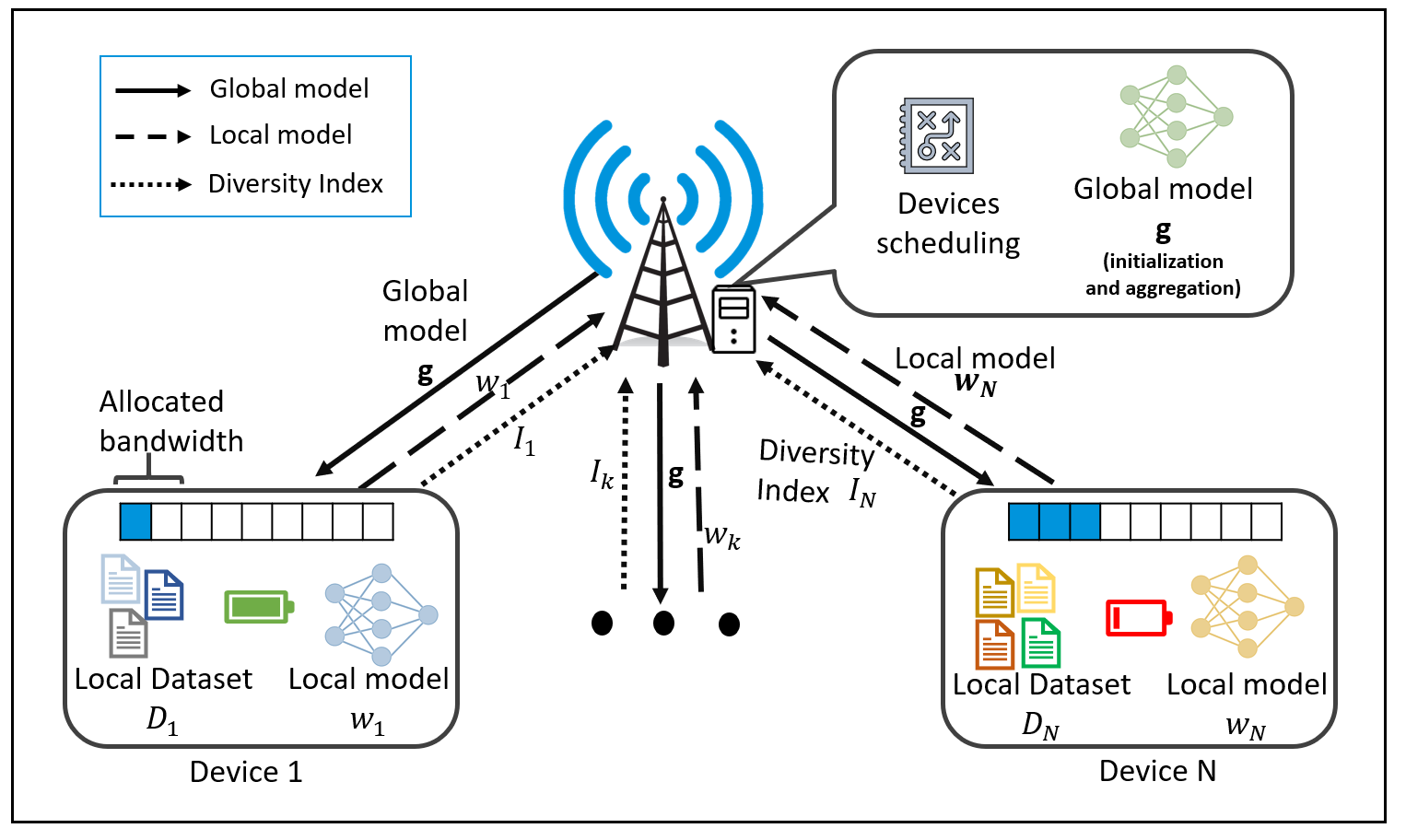}
	\caption{The proposed FEEL system model}
	\label{fig:fig_sys-model}
\end{figure*}

\begin{figure*}
	\centering
	\includegraphics[scale=0.64]{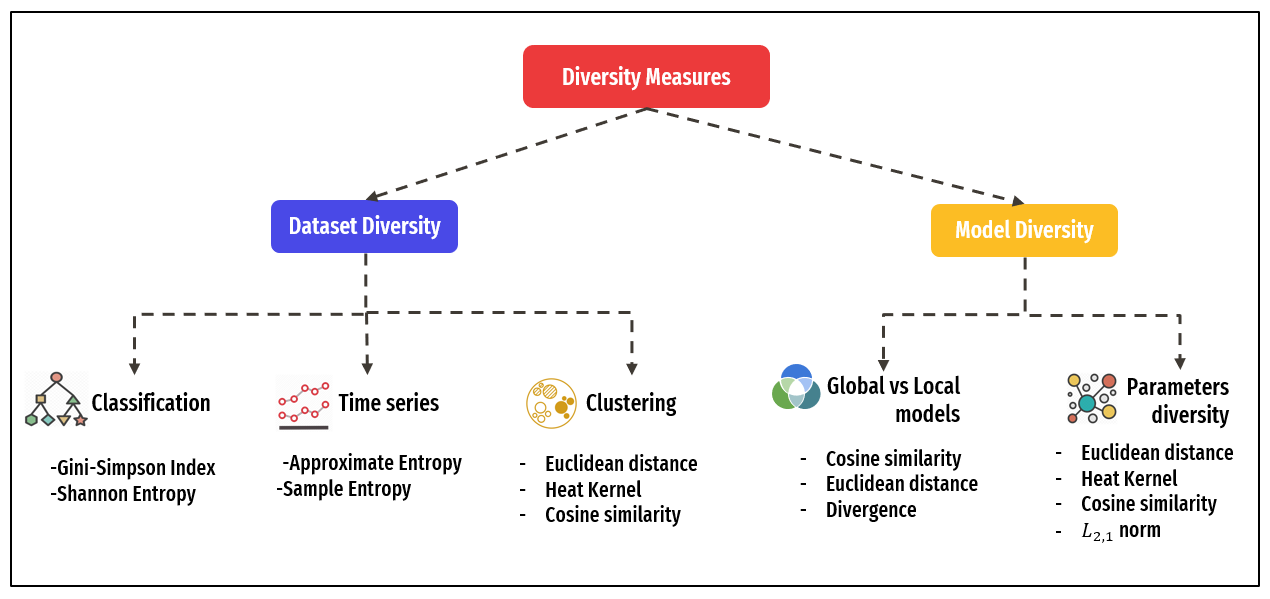}
	\caption{Diversity measures that can be used in pre-training and post-training scheduling}
	\label{fig:divmeasure}
\end{figure*}

Even if FL was first proposed with data as a central aspect, it has been overlooked in the design of proposed FEEL scheduling algorithms. With the significant drop of accuracy of models trained with resource-aware FEEL algorithms in non-IID and unbalanced settings, it becomes clear that the data aspect should be considered.
Henceforth, we propose a new possible data-aware end-to-end FEEL solution based on the diversity properties of the different datasets. In general, diversity consists of two aspects, namely, richness and uncertainty. Richness quantifies the size of the data, while the uncertainty quantifies the information contained in the data.  In fact, it has been long proven in Active Learning that by choosing highly uncertain data samples, a model can be trained using fewer labelled data samples. This fact suggests that data uncertainty should be incorporated into the design of FL scheduling algorithms. Nonetheless, the uncertainty measures used in Active Learning targets individual samples from unlabeled data in a centralized setting, thus, these measures cannot be directly integrated in FEEL.
In the FEEL setting, the updates' scheduling can be either before the training or after it, therefore the diversity measures should be selected depending on the time of scheduling.  If the scheduling before the training is preferred, then the datasets' diversity is to be considered. Otherwise, if the scheduling is set after the training is over, the diversity to be considered is model diversity, as the diversity of the dataset can be reflected by the resulting model. In both cases, in addition to maximizing the diversity through careful selection of participating devices, the scheduling algorithm can focus on minimizing the consumed resources in terms of completion time of FL and transmission energy of participating devices. For the pre-training scheduling, local computation energy can also be optimized. Furthermore, the scheduling problems' constraints are to be derived from the environment's properties concerning resources and data. \\
In this section, and to better illustrate the data-aware solutions, we consider the architecture illustrated in Figure \ref{fig:fig_sys-model}. The architecture is a cellular network composed of one base station (BS) equipped with a parameter server, and $\it{N}$ devices that collaboratively train a shared model.  
In the following, we discuss different constraints related to the scheduling algorithms in this setting. Then, we present pre-training and post-training algorithms guidelines, where we detail the key criteria for the design of data-aware FEEL solutions, and we present some potential measures and methods to enable a variety of data-aware FEEL applications, which are summarized in Figure \ref{fig:divmeasure}.

\subsection{Scheduling Constraints}
The scheduling algorithms' must consider the following constraints that arise from the FEEL environment's properties:
\textit{Energy consumption:} 
Due to the limited energy level and the high computational requirements of training algorithms, it is necessary to evaluate a device's battery level before scheduling it for a training round. 
When first FL was proposed, the selected devices were limited to the ones plugged for charging. However, this criterion limits the number of devices that can be selected, leading to a slow convergence of the learning.

\textit{Radio Channel State:} 
It is important to consider the radio channel state changes in the scheduling. The quality of the communication is critical for both the device selection and resource allocation. 

\textit{Expected completion time:}  
The available computation resources, alongside data size, can be used to estimate the completion time of the device. Potential stragglers can be discarded even before the training process. 

\textit{Number of participants:} A communication round cannot be considered valid unless a minimum number of updates is obtained. Therefore, a training round can be dropped if there are not enough devices to schedule.   

\textit{Data size:} The available data in the device is smaller in size than a required minimum, it can be immediately discarded from the selection process. For instance, if the number of samples is less than the selected mini-batch size, the device should be excluded.  

\subsection{Pre-training scheduling: Dataset Diversity}

The pre-training scheduling that we propose uses dataset diversity to choose devices that will conduct the training and send the updates. Scheduling the devices before the training allows to eliminate potential stragglers, and adapt the number of epochs based on the battery levels available at the participating devices.\\ 
\newline
\textit{1) Scheduling algorithm:}
\newline
In this algorithm, the global model is initialized by the BS. Afterwards, the following steps are repeated until the model converges or a maximum of rounds is attained: 
\begin{itemize}
\item \textbf{Step 1:}  At the beginning of each training round, the devices send their diversity indicators and battery levels to the server. 

\item  \textbf{Step 2:} Based on the received information, alongside with the evaluated channel state indicator, the server schedules a subset of devices and sends them the current global model. 

\item \textbf{Step 3:} Each device in the subset uses its local data to train the model. 

\item \textbf{Step 4:} The updated models are sent to the server to be aggregated.  

\item \textbf{Step 5:} The PS aggregates the updates and created the new model. 
\end{itemize}

\textit{2) Datasets Diversity Measures:}
\newline
In the pre-training scheduling, dataset diversity will serve essentially as a lead for device selection, where it should prioritize devices that have potentially informative datasets with less redundancy, to speed up the learning process.
While the richness of datasets can be easily quantified through the total number of samples, the uncertainty of the dataset depends strongly on the application.
For supervised learning, the uncertainty can be evaluated through the evenness of the dataset (i.e., the degree of balance between the classes in classification problems), which can be calculated through entropy measures. For sequence data, the uncertainty is reflected by the regularity of the series.  Moreover, for  unsupervised  learning, local dissimilarity between pseudo-classes or randomly sampled data points can be considered. 
Furthermore, it is essential to consider the privacy as a component of the used index. Sending the number of samples from each class for instance is a violation of the privacy principle of FEEL. In the following, we introduce some potential methods to evaluate datasets diversity.

\textbf{Diversity measures for classification:}
The measures of diversity have long been used in Active learning. In fact, uncertainty is used to choose the samples that should be labeled as this task is costly. However, in FL, the client selection does not concern independent samples, instead the diversity should be evaluated at the level of the entire dataset. 
Moreover, in the premise of supervised FL, the labels are already known, which gives the possibility to use more informed measures. For instance, Shannon Entropy or Gini-Simpson index are suitable measures for datasets' uncertainty in classification problems. 
Shannon Entropy and Gini-Simpson index both favor IID partitions, where the maximum for both indexes is obtained for balanced distributions and the datasets with a single class has the minimum possible value. 
The Shannon entropy quantifies the uncertainty (entropy or degree of surprise) of a prediction. It was first proposed to quantify the information content in strings of text. The underlying idea is that when a text contains more different letters, with almost equal proportional abundances, it will be more difficult to correctly predict which letter will be the next one in the string. However, Shannon Entropy is not defined for the case of classes with no representative samples. Therefore, it may not practical in scenarios with high unbalance. 
Another possible measure is the Gini-Simpson index. The Simpson index $\lambda$ measures the probability that two samples taken at random from the dataset of interest are from the same class. The Gini–Simpson index is its transformation $1-\lambda$, which represents the probability that the two samples belong to different classes. Nonetheless, if the number of classes is large, the distinction using this index will be hard. 

\textbf{Diversity measures for time-series forecasting:}
In time series problems, other methods can be used, such as Approximate Entropy (ApEn) and Sample Entropy (SampEn). In sequential data, statistical measures such as the mean and the variance are not enough to illustrate the regularity, as  they are influenced by system noise. ApEn is proposed to quantify the amount of regularity and the unpredictability of time-series data. It is based on the comparison between values of data in successive vectors, by quantifying how many data points vary more than a defined threshold. SampEn was proposed as a modification of  ApEn. It is used for assessing the complexity of time-series data, with the advantage of being independent from the length of the vectors.

\textbf{Diversity measures for clustering tasks:}
For clustering tasks, a similarity measure between data points from a randomly sampled subset should be considered. The measure can be distance based (e.g., Euclidean distance, Heat Kernel) or angular based (e.g., cosine similarity). A higher value is obtained if most of the data points in the sample are dissimilar, and thus the dataset should be considered as more diverse. It should be noted that angular based measures are invariant to scale, translation, rotation, and orientation, which makes them suitable for a wide range of applications, particularly multivariate datasets. \\

\subsection{Post-training scheduling: Model Diversity}

The post-training setting uses model diversity to choose devices that will send the updates. The model diversity is evaluated on two different aspects: 1) by comparing the dissimilarity between the local model's parameters and the previous global model's parameters. 2) by comparing the diversity within the model's parameters. 
In fact, choosing the local models that are divergent from the previous global model will possibly improve the representational ability of  the  global model directly, by aggregating updates that have potentially new information.  
Furthermore, if a dataset is highly  unbalanced and limited in size, the model's parameters would be very similar. The redundancy within parameters negatively  affects  the  model’s  representational ability. It is therefore necessary to prioritize updates with high diversity. 
In the following, we detail the post-training scheduling algorithm, then we present some possible measures for model diversity.   \\
\newline
\textit{1) Scheduling algorithm:}
\newline
Similarly to pre-training scheduling, the global model is initialized by the BS. Afterwards, the following steps are repeated until the model converges or a maximum of rounds is attained: 
\begin{itemize} 
\item  \textbf{Step 1:}  At the beginning of each training round, devices receive the current model. 

\item  \textbf{Step 2:} Each device in the subset uses its local data to train the model. 

\item  \textbf{Step 3:} The server sends an update request to the devices, to which each device responds by sending its model diversity index.

\item \textbf{Step 4:} Based on the received information, alongside with the evaluated channel state indicator, the server schedules a subset of devices to upload their models. Then, the updated models are sent to the server to be aggregated.  

\item  \textbf{Step 5:} The PS aggregates the updates and created the new global model. 
\end{itemize}

\textit{2) Model Diversity Measures:}
\newline
While the richness aspect of the diversity is irrelevant in models diversity due to fixed model size among devices, the information contained in the models can be quantified through how the local model's vary compared to the global model, and how the parameters within the same model repulse from each other. Some possible measures are as follows:

\textbf{Local and global models' dissimilarity:} 
Choosing the local models that are divergent from the previous global model will possible improve the representational ability of the  global model  directly \cite{wang_cmfl_2019}. Pairwise similarity measures such as cosine similarity and Euclidean distance can be used to evaluate the similarity of the new local parameters and the global parameters. Moreover, Divergence, a Bayesian method used to measure the difference between different data distributions, can be used to evaluate diversity of the learned model compared to the global model. 
Nonetheless, relying on model's dissimilarity might lead to collecting updates from outliers. It is thereby necessary to regulate these diversity measures through the use of thresholds in particular. 

\textbf{Parameters Dissimilarity:}
To evaluate the redundancy within the model's parameters, the same similarity measures used for clustering can also be applied to the parameters. 
Additionally, the $L_{2,1}$ norm can be used to obtain a group-wise sparse representation of the dissimilarity \cite{gong_diversity_2019}. The internal $L_1$ norm encourages different parameters to be sparse, while the external $L_2$ norm is used to control the complexity of entire model.





\section{Conclusion}
\label{sec:conclusion}
Federated Learning is a promising machine learning technique by virtue of its privacy-preserving aspect and ability to handle unbalanced and non-IID data. However, deploying federated learning based solutions at the edge of the network is subject to several challenges. In fact,
FEEL is a  multi-disciplinary problem that requires optimization over both the resources and the data. Nonetheless, the data properties are overlooked in many parts of the proposed algorithms, despite being the essence of federated learning.     
Several FEEL design challenges and issues are introduced and discussed in terms of trade-offs. Furthermore, a new research direction is presented in an effort to incorporate the datasets' diversity properties into the design of FEEL algorithms. Our proposed method supposes that the data quality and veracity are guaranteed, which requires leveraging other techniques such as the blockchain as a trusted third-party for data verification.

\section*{ACKNOWLEDGEMENT}
\label{sec:acknowledgement}
The authors would like to thank the Natural Sciences and Engineering Research Council of Canada, for the financial support of this research.
\bibliographystyle{IEEEtranN}
\bibliography{./includes/references}


\end{document}